# Droplet migration characteristics in confined oscillatory microflows

Kaustav Chaudhury, Shubhadeep Mandal, and Suman Chakraborty[*]

Department of Mechanical Engineering, Indian Institute of Technology Kharagpur

Kharagpur-721302, India

---

[*] Corresponding author, e-mail: suman@mech.iitkgp.ernet.in

# Abstract

We analyze the migration characteristics of a droplet in an oscillatory flow field in a parallel plate micro-confinement. Using phase filed formalism, we capture the dynamical evolution of the droplet over a wide range of the frequency of the imposed oscillation in the flow field, drop size relative to the channel gap, and the capillary number. The latter two factors imply the contribution of droplet deformability, commonly considered in the study of droplet migration under steady shear flow conditions. We show that the imposed oscillation brings in additional time complexity in the droplet movement, realized through temporally varying drop-shape, flow direction and the inertial response of the droplet. As a consequence, we observe a spatially complicated pathway of the droplet along the transverse direction, in sharp contrast to the smooth migration under a similar yet steady shear flow condition. Intuitively, the longitudinal component of the droplet movement is in tandem with the flow continuity and evolves with time at the same frequency as that of the imposed oscillation, although, with an amplitude decreasing with the frequency. The time complexity of the transverse component of the movement pattern, however, cannot by rationalized through such intuitive arguments. Towards bringing out the underlying physics, we further endeavor in a reciprocal identity based analysis. Following this approach, we unveil the time complexities of the droplet movement, which appear to be sufficient to rationalize the complex movement patterns observed through the comprehensive simulation studies. These results can be of profound importance in designing droplet based microfluidic systems in an oscillatory flow environment.

# I. INTRODUCTION

Droplet based microfluidics has brought in a revolution in modern science and technology, with applications ranging from material transport, active and passive mixing in smart and compact technologies, to cutting edge biophysical applications such as photo-initiated catalyst-initiated polymerization inside droplets, targeted drug delivery, and drug discovery [1–5]. In spite of the apparent diversities in many of these applications, the fundamental scientific challenge often boils down to a comprehensive assessment of the migration characteristics of the droplet in a micro-confined flow environment. While the migratory pattern of a droplet in presence of a steady background (extensional, simple shear, Poiseuille, etc.) flow has been studied quite extensively [6–16], investigations on the possible implications of a time-complex flow field in a confined microfluidic environment on the same are relatively scarce.

In contrast to the cases of a steady flow, a time complex flow field introduces an inertial contribution in the flow environment in addition to the time complex forcing [17]. The latter is commonly perceived as a time varying imposed pressure gradient. Subsequently, the flow field and the droplet shape both evolve with time along with a time varying inertial response of the droplet. Several studies have been reported in the literature on the dynamics of a droplet in time periodic flow environments [18–24]. However, those studies are mostly concerned with the discrete features in isolation: shape dynamics or hydrodynamic forces or streamwise migration characteristics. In contrast, combined confluence of the above mentioned factors needs to be taken under consideration for actual representation of the dynamics of the immersed droplet, particularly for lateral migration characteristics.

Here we present a comprehensive description of the migration characteristics of a droplet in an oscillating flow field, confined between two parallel plates. We employ both numerical simulations based on phase-field formalism, and concurrent analytical description based on reciprocal identity based paradigm towards developing the consensus. The present analysis, based on imposed oscillating flow, eventually unveils the essential intricacies of the dynamics of a confined droplet in a time complex flow field in general, as harmonic oscillation is an essential building block of any time complexity. Moreover, droplet dynamics within a confined flow environment is a representative of a class of soft-body dynamics within a confined physiological environment [25–27]. Therefore, the present study bears a potential impact on the dynamical features of biological moieties (such as red blood cells, while blood cells, etc.) within a physiological system where the true nature of the flow field is eventually time complex.

Even under the simple oscillating condition, we note a helical type movement of the droplet towards the lateral direction, in contrast to the typical smooth lateral migration of the droplet under steady shear. However, we find that the lateral migration complexities exhibit no particular biasness towards the droplet deformation.

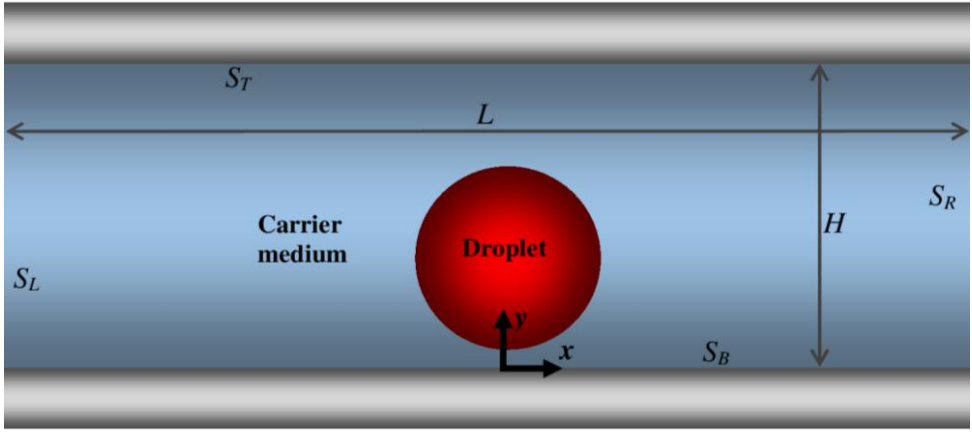

FIG.1. (Color online) Model problem setup, showing a Newtonian droplet immersed into another density and viscosity matched immiscible Newtonian carrier fluid, within a parallel plate channel.

Thus, implication of deformation based concept with inertia based corrections, as commonly employed for steady shear based analyses, falls short in describing the essential fact. Taking combined confluence of inertia and deformation along with flow alteration as essential factors, our theoretical paradigm explains the genesis of the observed time complex motion dynamics of the droplet, in satisfactory agreement with our simulation results.

## II. MODEL PROBLEM

We consider a Newtonian droplet of radius $a$, with dynamic viscosity $\eta$ and density $\rho$, initially submerged into another viscosity and density matched immiscible Newtonian fluid within a parallel plate channel, as shown in Fig. 1. The droplet is placed at certain distance offset from the channel centerline. The flow within the domain is then imposed by providing an oscillating body force. We carry out the analysis with respect to a Cartesian reference frame fixed to the channel wall, as shown in Fig. 1. Accordingly, the unit vectors are denoted by $\hat{\mathbf{e}}_i$ with $i$ being the corresponding coordinate. The imposed oscillating body force is acting along the $x$ direction.

### A. Phase field approach

We use the phase field model for simulating the dynamics of the two-fluid system. Particulars regarding this method are detailed elsewhere [28–32]. Here we briefly revise the features essential for our study. The distributions of the constituting fluid components are estimated by the distribution of an order parameter $\phi(\mathbf{x},t)$ (with $\mathbf{x}$ and $t$ being position vector and time respectively), such that, $\phi = -1$ denotes the droplet region and $\phi = +1$ denotes the carrier fluid. The interfacial region, between the droplet and the carrier fluid, is given by a diffused zone marked with $-1 < \phi < +1$.

Distribution of $\phi(\mathbf{x},t)$ leads to the free energy distribution $\mathscr{F}(\phi) = \int_\Omega \{f(\phi) + 2^{-1}\sigma\zeta|\nabla\phi|^2\}d\Omega$ over the entire volume of the domain $\Omega$. Here $f(\phi) = 4^{-1}\zeta^{-1}\sigma(\phi^2-1)^2$ represents the bulk free energy density exhibiting minima at the two stable components $\phi = \pm 1$. The $|\nabla\phi|^2$ term inside the integral represents the gradient free energy density, realized over the interfacial zone only. Here $\sigma$ and $\zeta$ denote the coefficient of surface tension and the characteristics thickness of the diffuse interfacial region respectively.

The energy principle requires $d\mathscr{F}/dt \leq 0$ leading to $(\delta\mathscr{F}/\delta\phi)(d\phi/dt) \leq 0$. However, the participating phases for the present situation should be conserved $d\phi/dt = -\nabla\cdot\mathbf{j}$ with $\mathbf{j}$ being the flux vector, for the transport of order parameter. These considerations lead to the celebrated Cahn-Hilliard equation [28–32]

$$\frac{\partial \phi}{\partial t} + \mathbf{u}\cdot\nabla\phi = \nabla\cdot(M\nabla\mu). \tag{1}$$

Here $M = M_c(1-\phi^2)$ denotes the inter phase mobility factor with $M_c$ as the critical mobility parameter, and $\mu = \delta\mathscr{F}/\delta\phi = \sigma\zeta^{-1}(\phi^3-\phi) - \sigma\zeta\nabla^2\phi$ is the chemical potential.

The velocity distribution $\mathbf{u}$ and subsequent advection of the order parameter (term $\mathbf{u}\cdot\nabla\phi$ in Eq. (1)) can be obtained from solving the continuity equation

$$\nabla\cdot\mathbf{u} = 0, \tag{2}$$

along with the Navier-Stokes equation

$$\frac{\partial}{\partial t}(\rho\mathbf{u}) + \nabla\cdot(\rho\mathbf{u}\mathbf{u}) = -\nabla p + \nabla\cdot\left[\eta\left(\nabla\mathbf{u}+\nabla\mathbf{u}^T\right)\right] - \nabla\cdot\mathbf{T}_S + G_0\sin(\omega t)\hat{\mathbf{e}}_x, \tag{3}$$

with $p$ as the pressure field. The stress tensor $\mathbf{T}_S$ is due to the interaction between the participating phases [28–30,33]. It is defined as $\mathbf{T}_S = [\partial\mathscr{L}/\partial(\nabla\phi)]\nabla\phi - \mathbf{I}\mathscr{L}$ with $\mathbf{I}$ being the identity tensor and the Lagrangian $\mathscr{L} = f(\phi) + \sigma\zeta(\nabla\phi)^2/2$ directly follows from the definition of the free energy functional $\mathscr{F}$ and satisfying the Euler-Lagrange equation. For the present consideration, the force of interaction between the participating components eventually simplifies to $-\nabla \cdot \mathbf{T}_S = \mu \nabla \phi$ [28–30,33]. The body force $G_0 \sin(\omega t)\hat{\mathbf{e}}_x$ in Eq. (3) delineates the forcing required to maintain the oscillation in the flow field, with $G_0$ and $\omega$ being the amplitude and the frequency, respectively, of the imposed forcing.

Governing Eqs. (1)-(3) are corroborated with the following set of boundary conditions (please refer Fig. 1 for reference):

At $S_T$ and $S_B$ (with $\hat{\mathbf{n}}_s$ denoting the unit vector normal to a solid surface)

$$\begin{aligned}&(a)\, \mathbf{u}-(\mathbf{u}\cdot\hat{\mathbf{n}}_s)\hat{\mathbf{n}}_s = 0 \quad \text{(no slip)}\\&(b)\, \mathbf{u}\cdot\hat{\mathbf{n}}_s = 0 \quad \text{(no penetration)}\\&(c)\, \hat{\mathbf{n}}_s\cdot\nabla\phi = 0 \ \text{ and }\ \hat{\mathbf{n}}_s\cdot\nabla\mu = 0 \quad \text{(no flux)}\end{aligned} \qquad (4)$$

and periodic over $S_L$ and $S_R$, so that

$$\mathbf{u}(\mathbf{x})=\mathbf{u}(\mathbf{x}+L),\ p(\mathbf{x})=p(\mathbf{x}+L),\ \text{and}\ \phi(\mathbf{x})=\phi(\mathbf{x}+L). \qquad (5)$$

### B. Nondimensional parameters

We perform numerical experimentations over the parametric space of the following set of nondimensional parameters: Péclet number $\text{Pe} = H^2 U_c / M_c \sigma$, Reynolds number $\text{Re} = \rho U_c H / \eta$, capillary number $\text{Ca} = \eta U_c / \sigma$, Strouhal number $\text{St} = \omega H / U_c$, Cahn number $\text{Cn} = \zeta / H$ and the relative drop size $a/H$. The Characteristic velocity scale is chosen as $U_c = G_0 H / 8\eta^2$.

Here Pe denotes the relative importance of the advection of $\phi$ over the diffusion, as defined through Cahn-Hilliard Eq. (1). The appearance of $\sigma$ in Pe is due to the definition of the free energy functional $\mathscr{F}$ and chemical potential $\mu$ using $\zeta$ and $\sigma$. This is attributed to the connection between the surface tension with the excess free energy across the interface [28–32]. Among all the parameters discussed above, Cn and Pe are guided by the numerical constraints, to obtain feasible solution, sufficient to capture the essential features of interest [34,35]. Regarding the remaining parameters, we deliberately keep Re small, so as to mimic common microfluidic applications. We primarily focus on the deformation and the subsequent dynamical features of the droplet, associated with the oscillation in the flow field. Therefore, we vary St (signifying the nondimensional frequency), in addition to Ca and $a/H$.

## C. Implementation

In reference to Fig. 1, we select a rectangular domain of size (normalized by channel height $H$) $3\times 1$ which is tessellated with uniform grids. The transport equations are then discretized invoking the finite volume formalism along with semi-implicit time discretization policy. For estimation of the flux functions across the faces of the control volume, we consider second order upwind scheme. For evaluating all other terms requiring calculation of the values at the faces of the control volumes, central differencing scheme is employed in the present study. We consider SIMPLE [36] strategy for pressure-velocity coupling. The discretized equations are then solved using algebraic multi-grid (AMG) method [37].

In presenting the results, we consider the migration characteristics given by the droplet position $(x_d, y_d)$ (both normalized by $H$) and the droplet velocity $(u_{dx}, u_{dy})$ (both normalized by $U_c$), as measured along $x$ and $y$ directions, respectively. Hereinafter whenever we refer to the mentioned parameters, it should be considered as the normalized versions, unless otherwise specified. Time $t$ will also be referred as its normalized version scaled with $H/U_c$.

## III. RESULTS

### A. Plane Poiseuille flow condition and comparison with reported literature

In the low Re regime, the shape of a droplet in a plane Poiseuille flow at a given lateral location depends on $a/H$ and Ca, keeping other parameters fixed [38–41,11]. Theoretically, in the limit $\text{Ca} \ll 1$ and $a/H \ll 1$, the drop shape can be given by

$$\frac{r}{a} = 1 + \text{Ca}\left[ 4\left(1-\frac{y_d}{H}\right)\frac{16+9\lambda}{8+8\lambda}\frac{x'y'}{r^2}\left(\frac{a}{H}\right) \right.$$
$$\left. +4\frac{10+11\lambda}{40+40\lambda}\left(\frac{x'}{r}-\frac{5x'y'^2}{r^3}\right)\left(\frac{a}{H}\right)^2 + O\left(\left(\frac{a}{H}\right)^3\right) \right] + O(\text{Ca}^2) \quad (6)$$

where $x' = x - x_d$, $y' = y - y_d$, and $\lambda$ denotes the viscosity contrast, defined as the ratio of the dynamic viscosity of the droplet to that of the carrier medium. We restrict our attention to the case $\lambda = 1$. Evidently, the approximation $r/a = 1$ from Eq. (6) is reminiscent of the leading order spherical shape in a low Reynolds number flow. The remaining terms in Eq. (6) demonstrate the deviation from sphericity. The deformation is characterized by the nondimensional numbers Ca and $a/H$. The former denotes the deformation due to viscous forcing in comparison to the shape preserving capillary forcing. The latter, on the other hand, signifies the contribution of the bounding walls.

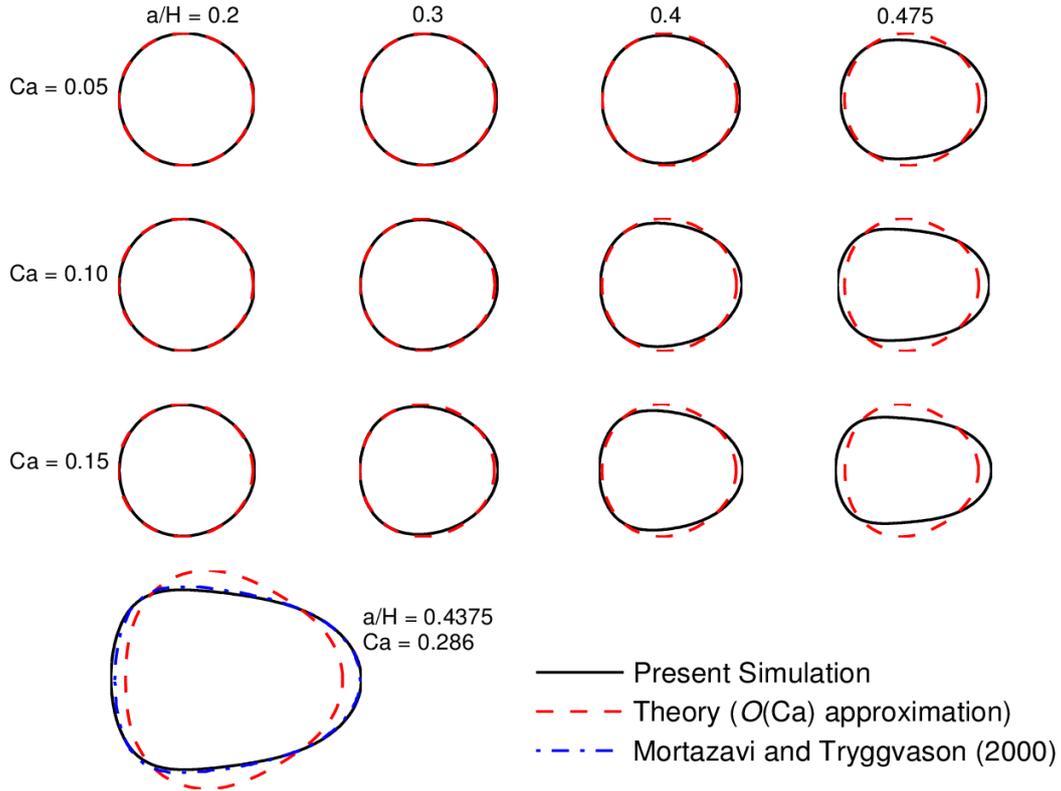

FIG. 2. (Color online) Characteristic shapes of droplets at different $a/H$ and Ca in plane Poiseuille flow condition, as obtained from the present simulations. For small $a/H$ and Ca, the comparisons are made against the $O(\text{Ca})$ theoretical approximation (Eq. (6)). For larger droplet $(a/H = 0.4375)$ at higher $\text{Ca}(= 0.286)$, the present simulation result is compared against the simulation result of Mortazavi and Tryggvason [15]. Note that here the shapes are shown at 1:1 aspect ratio with axis-tight format. This makes droplets of all sizes to appear as almost same size, thereby allowing better visualization of the droplets with small $a/H$. However, the actual sizes of the droplets are in accordance with the respective $a/H$.

In Fig. 2 we compare our simulation prediction of drop shape with the theoretical estimation through Eq. (6) and with the numerical results of Mortazavi and Tryggvason [15]. From the figure it is evident that a close agreement with the theoretical estimation in the small $a/H$ and Ca regime is obtained with our simulation setup for grid size (normalized) $\Delta x = 0.0075$ in both $x$ and $y$ directions with $\text{Cn} = 0.015$ and $\text{Pe} = 3 \times 10^6$. At higher $a/H$, the wall induced deformation produces highly elongated droplet which deviates much from the theoretical approximation in Eq. (6). For larger drops, we find that $\Delta x = 0.01$ with $\text{Cn} = 0.02$ and $\text{Pe} = 10^6$ is sufficient to capture the droplet dynamics, taking both accuracy and computational expenses into consideration. The setup yields a satisfactorily matching with the shape prediction in Ref. [15], as evident from Fig. 2. These predictions are also in qualitative agreement with experimental findings.

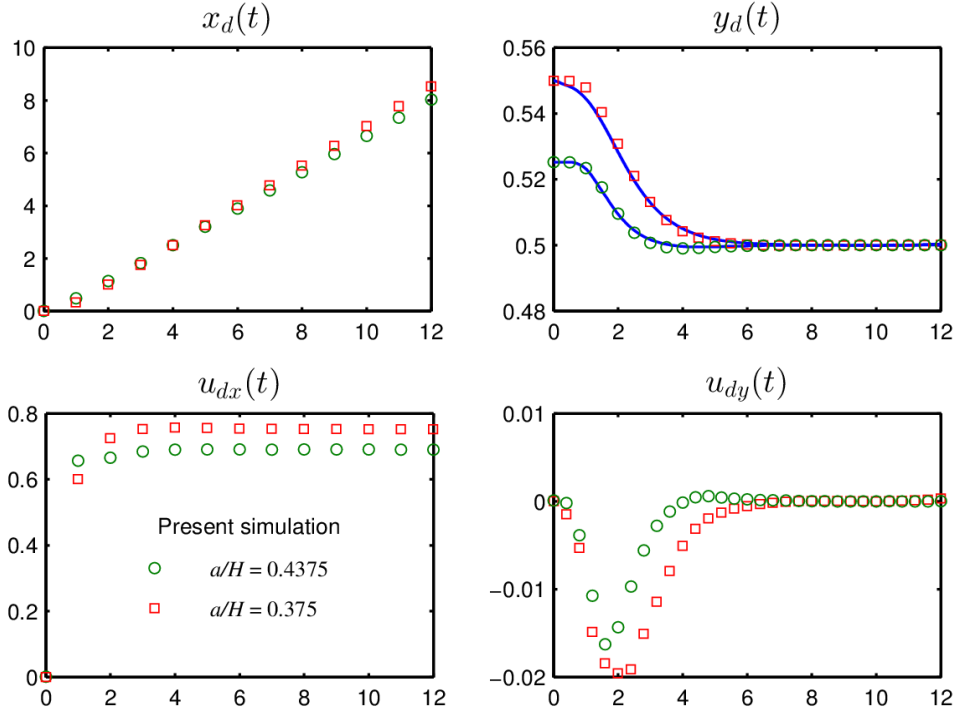

FIG. 3. (Color online) The time course of the migration characteristics of a droplet in a plane Poiseuille flow condition, for $a/H = 0.4375$ and $0.375$ (indicated by the open markers), with Re = 1 and 4 respectively, and Ca = 0.286. The parametric setup is in tune with the setup of Mortazavi and Tryggvason [15]; their results are indicated by the corresponding solid lines.

In Fig. 3, we present the temporal evolution of the droplet position $x_d(t)$, $y_d(t)$ and the corresponding migration speeds $u_{dx}(t)$ and $u_{dy}(t)$. Here the results are shown for $a/H = 0.4375$ and $0.375$, with Re = 1 and 4 respectively, and Ca = 0.286. This is in accordance with the setup in Ref. [15]. Essentially this setting gives us an idea about the dynamics of confined droplets. To obtain grid independent results, we have used $\Delta x = 0.01$ with Cn=0.02 and Pe=$10^6$. Starting from the initial position, a droplet always tends to assume a lateral position, from where onwards there will be no acceleration of the droplet, as evident from Fig. 3. This condition is the so called equilibrium settling of the droplet.

## B. Droplet migration under oscillatory flow condition

In Fig. 4, we show the trajectories of a droplet under oscillatory flow conditions at different $St$. For the sake of comparison, we also present the trajectory of the same droplet under a plane Poiseuille flow condition. From the figure, it appears that at lower $St$ regime, the droplet follows almost the similar path as can be observed under plane Poiseuille flow condition. However, a characteristic oscillatory motion parallel to the channel walls is noteworthy. Interestingly, at higher $St$ regime, the droplet follows a complex helical type pathway. This feature is in sharp contrast to the droplet trajectory under a plane Poiseuille flow condition. Following the helical type pathway, a net transverse displacement of the droplet, without any considerable displacement along $x$ direction, can be observed specifically in the higher $St$ regime. This is of immense significance in practical cases where one tries to move the droplet towards a particular transverse location, but is constrained by the axial extent of the channel.

To unveil the characteristics of the complex movement, it is imperative to consider the temporal evolution characteristics $x_d(t)$, $y_d(t)$, $u_{dx}(t)$ and $u_{dy}(t)$. Fig. 5 demonstrates those characteristics, for $Ca = 0.286$, $Re = 1$ and $a/H = 0.4375$. Here we present the characteristics with respect to $\theta = t\,St$, to have a common basis for analysis for different $St$. From Fig. 5 it appears that unlike the plane Poiseuille flow condition, here $x_d(t)$ and $u_{dx}(t)$ both follow oscillating nature. This behavior seems intuitive, and can be attributed to the change in momentum along $x$ direction due to the oscillating body force acting along the same direction. However, the oscillating pattern in the transverse migration characteristics $y_d(t)$ and $u_{dy}(t)$, as can be seen in Fig. 5, are difficult to rationalize by such simple argument.

From the $u_{dx}(\theta)$ variations in Fig. 5, it appears that $St$ has little impact on $u_{dx}$, at least in the low $St$ regime. However, in the $x_d(\theta)$ characteristics, we note a decreasing trend of the amplitude of $x_d$ with $St$. Additionally, in Fig. 5, we also observe the oscillating nature of the transverse migration characteristics $y_d(\theta)$ and $u_{dy}(\theta)$ at all $St$ values considered here. Nevertheless, a net lateral migration from the initial position is noteworthy.

In Fig. 6 we show the migration characteristics for $St = 1$ at different $Ca$ (Fig. 6a) and $a/H$ (Fig. 6b) values. Deformability of a droplet is considered to be one of the prime contributors to the cross-stream migration characteristics. In the small $a/H$ and $Ca$ limit, droplet deformation is expected to be small. From Fig. 6 it is evident that under such condition, transverse oscillation still persists. Thus, it appears that oscillation in the $y_d(t)$ and $u_{dy}(t)$ characteristics is an inherent feature that is associated with the oscillation in the imposed flow field.

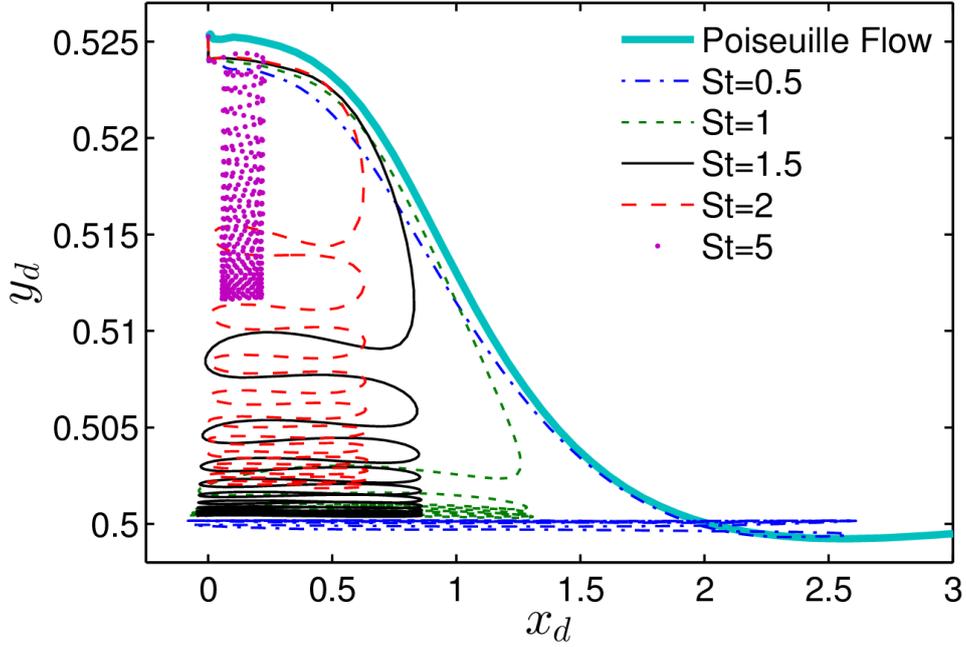

FIG. 4. (Color online) Trajectory of a confined droplet ($a/H = 0.4375$ and $Ca = 0.286$) in an oscillating flow field, for different $St$. The numerical setup corresponds to $Cn = 0.02$, $\Delta x = 0.01$ and $Pe = 10^6$, taking accuracy and computational expenses into combined consideration. In the figure the abscissa and ordinate represent the normalized *x* and *y* coordinates respectively.

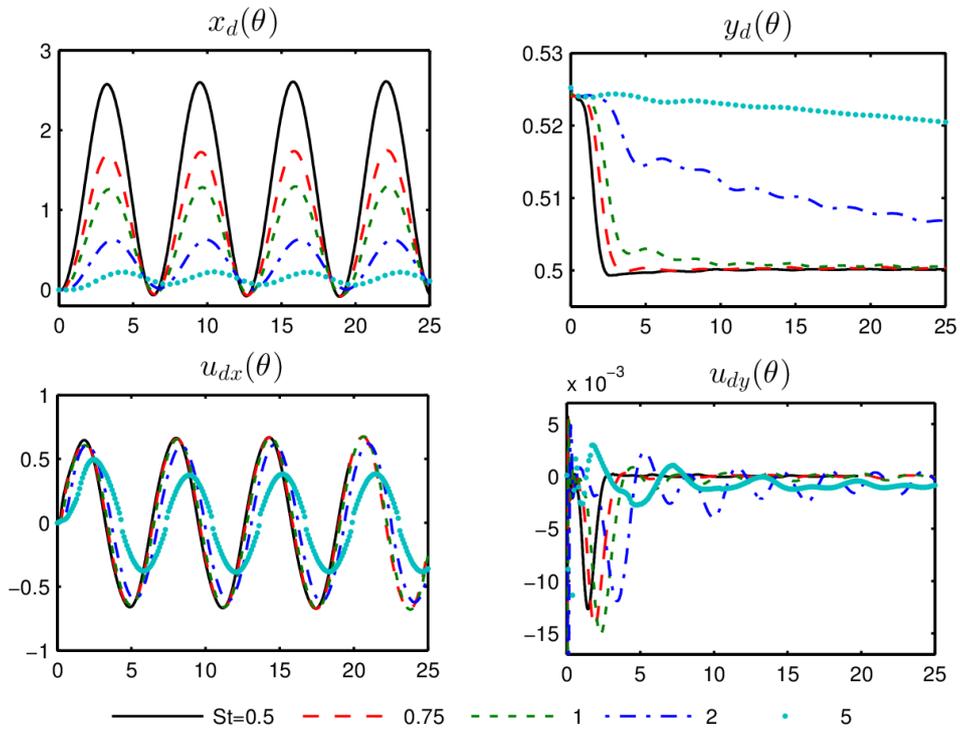

FIG. 5. (Color online) Characteristics of the pathway of a droplet with $a/H = 0.4375$, $Re = 1$ and $Ca = 0.286$, in an oscillating flow field at different $St$.

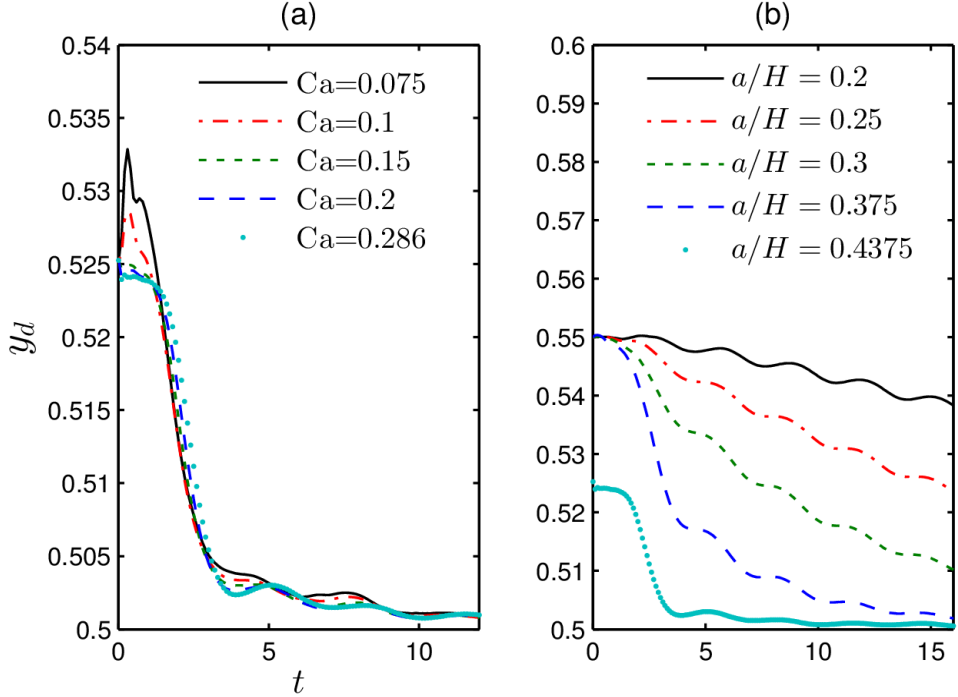

FIG. 6. (Color online) The migration characteristics of a droplet in an oscillating flow field, for St=1 at different (a) Ca (keeping $a/H = 0.4375$) and (b) $a/H$ (keeping Ca=0.286), with Re = 1 for both the cases.

## IV. AN ANALYTICAL DESCRIPTION FOR THE DROPLET PATHWAY

Here our primary goal is to unveil an estimation of the time complexity of $x_d(t)$, $y_d(t)$, $u_{dx}(t)$ and $u_{dy}(t)$ due to the imposed time complex flow field, leading towards the complex movement of a droplet within a confinement. To this end, we proceed with the reciprocal identity based analysis. Our problem can be generalized as the motion of a viscous droplet of dynamic viscosity $\lambda \eta$ and density $\rho$ in a unidirectional time complex flow field of an incompressible Newtonian fluid of dynamic viscosity $\eta$ and density $\rho$, in between two infinite parallel plates. Specifically, for the case studies presented in this work, $\lambda = 1$. However, we proceed with the analysis of arbitrary $\lambda$, for the sake of completeness.

For theoretical analysis, it is convenient to analyze the situation with respect to the droplet reference frame. However, the position $(x_d, y_d)$ and the velocity $(\mathbf{u}_d)$ (in vector form) of the droplet is measured with respect to the fixed $x - y$ reference frame. Now, we define a Cartesian reference frame $\xi - \psi$ fixed to the droplet where its origin coincides with the droplet centroid with $\xi$ and $\psi$ aligns parallel to $x$ and $y$ axes respectively. Hereinafter, the Cartesian reference frame $\xi - \psi$ will be referred as the *droplet reference frame*, unless otherwise specified, and the position vector with respect to this frame will be given by $\mathbf{r}$.

In our simulations, we consider that the oscillation in flow field is imposed by an oscillating body force. For theoretical convenience, here we assume this to be incorporated into the pressure gradient term for the imposed flow field. In accordance with our problem, we consider $\text{Re} = \rho U_c a/\eta \ll 1$. Note that here the characteristic length scale for normalization is $a$, instead of $H$ (considered for normalization of our simulation model). Other scales for normalization are same as in our simulation setup. Additionally, here we also consider $\text{Ca} = \eta U_c/\sigma \ll 1$. The small Ca approximation essentially allows analytical tractability, for droplet based problems [11]. Moreover, the theoretical understanding obtained from small Ca based analysis can also be extended to moderate and even sometimes high Ca cases, at least for unveiling the essential features of interest.

Pertaining to the present situation under consideration, the governing equations, in nondimensional form, with respect to the droplet reference frame can be recast as

$$(a)\nabla\cdot\mathbf{u} = 0 \text{ and } \nabla\cdot\mathbf{T} = \text{Re}\left(\frac{\partial \mathbf{u}}{\partial t} + \mathbf{u}\cdot\nabla\mathbf{u}\right) \text{ with } \mathbf{T} = -p\mathbf{I} + \left(\nabla\mathbf{u} + \nabla\mathbf{u}^T\right),$$

$$(b)\nabla\cdot\bar{\mathbf{u}} = 0 \text{ and } \nabla\cdot\bar{\mathbf{T}} = \text{Re}\left(\frac{\partial \bar{\mathbf{u}}}{\partial t} + \bar{\mathbf{u}}\cdot\nabla\bar{\mathbf{u}}\right) \text{ with } \bar{\mathbf{T}} = -\lambda\bar{p}\mathbf{I} + \lambda\left(\nabla\bar{\mathbf{u}} + \nabla\bar{\mathbf{u}}^T\right), \quad (7)$$

$$(c)\nabla\cdot\mathbf{u}_\infty = 0 \text{ and } \nabla\cdot\mathbf{T}_\infty = \text{Re}\left(\frac{\partial \mathbf{u}_\infty}{\partial t} + \mathbf{u}_\infty\cdot\nabla\mathbf{u}_\infty\right) \text{ with } \mathbf{T}_\infty = -p_\infty\mathbf{I} + \left(\nabla\mathbf{u}_\infty + \nabla\mathbf{u}_\infty^T\right)$$

for the region (a) outside and (b) inside the droplet, and for the (c) imposed flow field. The velocity, pressure and the stress distributions for the corresponding domains are denoted, respectively, by $(\mathbf{u}, p, \mathbf{T})$, $(\bar{\mathbf{u}}, \bar{p}, \bar{\mathbf{T}})$ and $(\mathbf{u}_\infty, p_\infty, \mathbf{T}_\infty)$. The above equations are corroborated with the boundary conditions

$$\begin{aligned}(a)&\,\mathbf{u} \to \mathbf{u}_\infty \text{ as } r \to \infty \quad \left(\text{with } r = |\mathbf{r}|\right),\\(b)&\,\mathbf{u} = -\mathbf{u}_d \text{ on the walls,}\end{aligned} \quad (8)$$

and interfacial conditions at the surface of the droplet

$$\begin{aligned}(a)&\,\mathbf{u} = \bar{\mathbf{u}},\\(b)&\,\mathbf{u}\cdot\hat{\mathbf{n}} = \bar{\mathbf{u}}\cdot\hat{\mathbf{n}} = 0,\\(c)&\,\mathbf{T}\cdot\hat{\mathbf{n}} - \bar{\mathbf{T}}\cdot\hat{\mathbf{n}} = \frac{1}{\text{Ca}}\kappa\hat{\mathbf{n}}.\end{aligned} \quad (9)$$

Here $\kappa$ denotes the local curvature of the surface of the droplet with $\hat{\mathbf{n}}$ being the unit outward normal.

## A. The reciprocal identity

Our theoretical estimation is centered on the reciprocal identity based analysis. In this method, the analysis commences with the consideration of a complementary problem such that the shape of the droplet for the complementary problem and the original problem are same. To this end, we consider the motion of a Newtonian droplet in a density and viscosity matched Newtonian fluid where the droplet is translating perpendicular to the walls of the channel. Accordingly, its flow characteristics outside and inside the droplet are given by $(\mathbf{u}_c, p_c, \mathbf{T}_c)$ and $(\bar{\mathbf{u}}_c, \bar{p}_c, \bar{\mathbf{T}}_c)$ respectively. The governing equations, along with the boundary and interfacial conditions, are having standard forms, and adopted here as it is [17,23,40,11]. Thus, we prefer not to repeat them again. We must note that the original problem can be reduced by means of domain perturbation to equivalent spherical droplet case, for applying the interfacial conditions [17,11]. Thus, we may conveniently proceed with the spherical droplet consideration for the complementary problem.

Following the standard procedure of reciprocal based analysis, we eventually end up into the identity of the form

$$\int_{S_d} \left[ (\mathbf{T} - \bar{\mathbf{T}}) \cdot \mathbf{u}_c - (\mathbf{T}_c - \bar{\mathbf{T}}_c) \cdot \mathbf{u} - \bar{\mathbf{T}}_c \cdot (\mathbf{u} - \bar{\mathbf{u}}) - \mathbf{T}_\infty \cdot \mathbf{u}_c + \mathbf{T}_c \cdot \mathbf{u}_\infty \right] \cdot \hat{\mathbf{e}}_r dS =$$

$$-\mathrm{Re} \left\{ \int_\Omega \left[ \left( \frac{\partial \mathbf{u}}{\partial t} + \mathbf{u} \cdot \nabla \mathbf{u} \right) - \left( \frac{\partial \mathbf{u}_\infty}{\partial t} + \mathbf{u}_\infty \cdot \nabla \mathbf{u}_\infty \right) \right] \cdot \mathbf{u}_c d\Omega \quad (10) \right.$$

$$\left. - \int_\Omega \left[ \frac{\partial \bar{\mathbf{u}}}{\partial t} + \bar{\mathbf{u}} \cdot \nabla \bar{\mathbf{u}} \right] \cdot \bar{\mathbf{u}}_c d\Omega \right\},$$

where $S_d$ denotes the surface of the droplet and $\Omega$ denotes the volume of the entire problem domain. The procedure to arrive at the reciprocal identity is similar to that outlined in Refs. [17,11]. However, here the only difference is the inclusion of the inertial volume integrals, attributable to the time complexity involved with the problem. Similar volume integrals can be observed in Refs. [22,23], dealing with the inertial contribution in droplet dynamics.

## B. Asymptotic expansion

Owing to the consideration of small Re and Ca, it is possible to proceed with the double asymptotic expansion [11]

$$A = A^{(0)} + \mathrm{Re}\, A^{(\mathrm{Re})} + \mathrm{Ca}\, A^{(\mathrm{Ca})} + \mathrm{Ca}^2 A^{(\mathrm{CaCa})}$$
$$+ \mathrm{Re}\,\mathrm{Ca}\, A^{(\mathrm{ReCa})} + \mathrm{Re}^2 A^{(\mathrm{ReRe})} + \ldots, \quad (11)$$

for any arbitrary function $A$ which can be anything from $\mathbf{u}_d$, $(\mathbf{u}, p, \mathbf{T})$, $(\bar{\mathbf{u}}, \bar{p}, \bar{\mathbf{T}})$ and $(\mathbf{u}_\infty, p_\infty, \mathbf{T}_\infty)$. For $\bar{p}$ and $\bar{\mathbf{T}}$, additionally, we need to consider the $O(\text{Ca}^{-1})$ contribution towards satisfying the Laplace pressure jump [11].

Following the above mentioned approach, the shape of the droplet can be approximated as

$$F \equiv r - 1 - \underbrace{\left(\text{Ca}\, f^{(\text{Ca})} + \text{Re}\,\text{Ca}\, f^{(\text{ReCa})} + \text{Ca}^2\, f^{(\text{CaCa})} + \ldots\right)}_{\mathscr{z}(\mathbf{r},t)} = 0 \quad (12)$$

where $\mathscr{z}$ signifies the deviation from sphericity expressed in terms of yet unknown function $f$. Note that in expanding $\mathscr{z}$, the first contribution of Re comes from the $O(\text{ReCa})$ term. Actually, the droplet can deform even in complete absence of inertia, as signified by the $O(\text{Ca})$ contribution. In presence of inertia, however, the deformation is contributed by the combined influence of Re and Ca. Using $F$, now we can define

$$\hat{\mathbf{n}} = \frac{\nabla F}{|\nabla F|} = \hat{\mathbf{e}}_r - \text{Ca}\,\nabla f^{(\text{Ca})} - \text{Re}\,\text{Ca}\,\nabla f^{(\text{ReCa})}$$
$$- \text{Ca}^2 \left[\nabla f^{(\text{CaCa})} + \frac{\left(\nabla f^{(\text{Ca})} \cdot \nabla f^{(\text{Ca})}\right)\hat{\mathbf{e}}_r}{2}\right] - \ldots, \quad (13)$$

and

$$\kappa = \nabla \cdot \hat{\mathbf{n}} = 2 - \text{Ca}\left[2 f^{(\text{Ca})} + \nabla^2 f^{(\text{Ca})}\right] - \text{Re}\,\text{Ca}\left[2 f^{(\text{ReCa})} + \nabla^2 f^{(\text{ReCa})}\right]$$
$$- \text{Ca}^2 \left[2 f^{(\text{CaCa})} - 2 f^{(\text{Ca})} f^{(\text{Ca})} + \nabla^2 f^{(\text{CaCa})}\right] - \ldots. \quad (14)$$

with $\hat{\mathbf{e}}_r$ as the unit vector along the radial direction.

### C. The case of spherical droplet: the $O(1)$ problem

The governing equations for the $O(1)$ problem are

$$\begin{aligned}(a)\,& \nabla \cdot \mathbf{u}^{(0)} = 0 \text{ and } \nabla \cdot \mathbf{T}^{(0)} = 0, \\ (b)\,& \nabla \cdot \bar{\mathbf{u}}^{(0)} = 0 \text{ and } \nabla \cdot \bar{\mathbf{T}}^{(0)} = 0, \\ (c)\,& \nabla \cdot \mathbf{u}_\infty^{(0)} = 0 \text{ and } \nabla \cdot \mathbf{T}_\infty^{(0)} = 0,\end{aligned} \quad (15)$$

with the boundary and interfacial conditions

$$\begin{aligned}&(i)\, \mathbf{u}^{(0)} \to \mathbf{u}_\infty^{(0)} \text{ as } r \to \infty,\\ &(ii)\, \mathbf{u}^{(0)} = -\mathbf{u}_d^{(0)} \text{ on the walls,}\\ &(iii)\, \mathbf{u}^{(0)} = \bar{\mathbf{u}}^{(0)} \text{ at } r = 1, \quad\quad\quad (16)\\ &(iv)\, \mathbf{u}^{(0)} \cdot \hat{\mathbf{e}}_r = \bar{\mathbf{u}}^{(0)} \cdot \hat{\mathbf{e}}_r = 0 \text{ at } r = 1, \text{ and}\\ &(v)\, \mathbf{T}^{(0)} \cdot \hat{\mathbf{e}}_r - \bar{\mathbf{T}}^{(0)} \cdot \hat{\mathbf{e}}_r = -\left[2 f^{(\text{Ca})} + \nabla^2 f^{(\text{Ca})}\right]\hat{\mathbf{e}}_r \text{ at } r = 1.\end{aligned}$$

Solution for the imposed flow field can be recast as

$$\mathbf{u}_\infty^{(0)} = \left(\alpha + \beta\psi + \gamma\psi^2\right)\Gamma(t)\hat{\mathbf{e}}_\xi - \mathbf{u}_d^{(0)} \quad\quad (17)$$

where $\alpha = 4 y_d (1 - y_d)$, $\beta = 4(1 - 2 y_d) a/H$ and $\gamma = -4(a/H)^2$. In Eq. (17) $\Gamma(t) = \varepsilon_1 + \varepsilon_2 \sin(t\,\text{St})$ such that $\varepsilon_1 = 1, \varepsilon_2 = 0$ represents Poiseuille flow condition and $\varepsilon_1 = 0, \varepsilon_2 = 1$ denotes oscillating flow condition. It is to be noted that for the spherical droplet problem, one can directly adopt the solution proposed by Chan and Leal [11], with the coefficients $\alpha$, $\beta$ and $\gamma$ of their paper be replaced by $\alpha\Gamma(t)$, $\beta\Gamma(t)$ and $\gamma\Gamma(t)$. We must appreciate that at the $O(1)$ approximation it is possible to estimate $u_{dx}^{(0)} = \mathbf{u}_d^{(0)} \cdot \hat{\mathbf{e}}_x = \mathbf{u}_d^{(0)} \cdot \hat{\mathbf{e}}_\xi$. Additionally, due to aft-fore symmetry, $u_{dy}^{(0)} = \mathbf{u}_d^{(0)} \cdot \hat{\mathbf{e}}_y = \mathbf{u}_d^{(0)} \cdot \hat{\mathbf{e}}_\psi = 0$. Now, proceeding with the notion that the net force acting on the neutrally buoyant droplet is zero, one can obtain

$$u_{dx}^{(0)} = \Lambda_{CL}\Gamma(t) \quad\quad (18)$$

where $\Lambda_{CL} = \alpha + \gamma\lambda/(2 + 3\lambda) + I_1 + I_4/2$ represents the same function as given in Ref. [11]. Now we proceed with the form of $\Gamma(t) = \varepsilon_1 + \varepsilon_2 \sin(t\,\text{St})$.

For Poiseuille flow condition $(\varepsilon_1 = 1, \varepsilon_2 = 0)$, Eq. (18) degenerates to the form $u_{dx}^{(0)} = \Lambda_{CL} = \alpha + \gamma\lambda/(2 + 3\lambda) + I_1 + I_4/2$. This is a very popular estimation of streamwise migration speed of a droplet in plane Poiseuille flow. In Sec III we have already presented the form $u_{dx} = \alpha + \gamma\lambda/(2 + 3\lambda) + O\left((a/H)^3\right)$, with the note that the $O\left((a/H)^3\right)$ contribution emerges from the consideration of wall effects. Using the method of reflection, theoretically it is possible to obtain the wall effects which provide the correction $I_1 + I_4/2 = O\left((a/H)^3\right)$ [11].

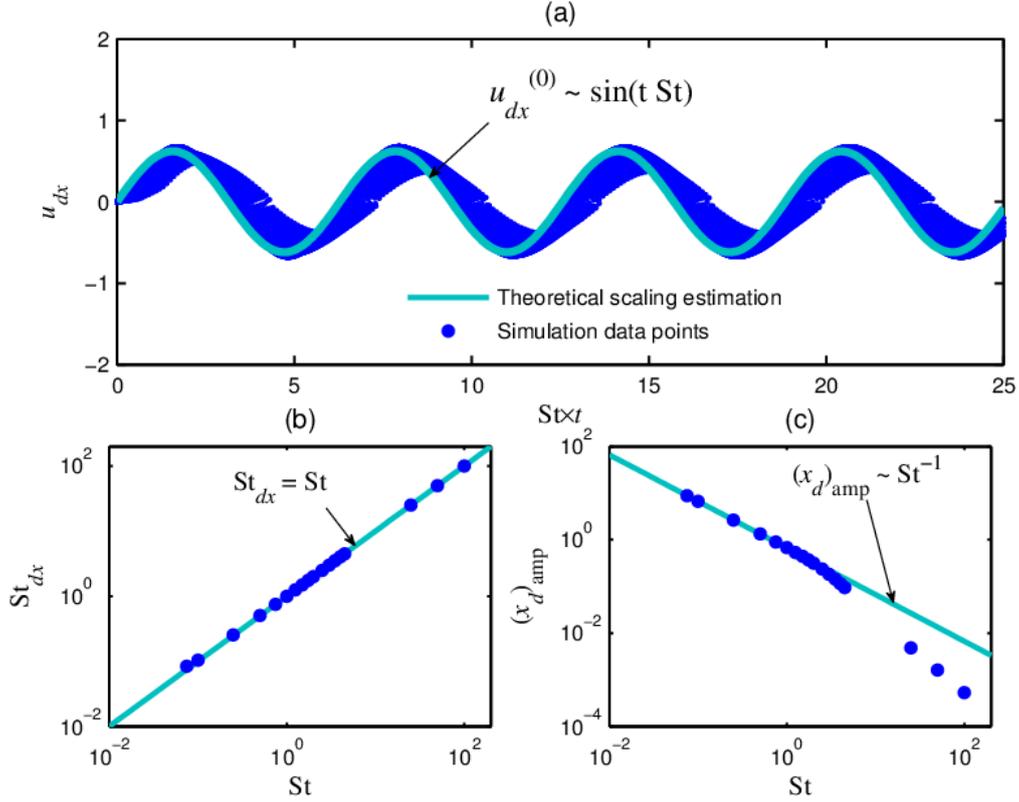

FIG. 7. (Color online) Comparison of the present simulation results on the streamwise migration characteristics with the $O(1)$ theoretical approximation. In the figure, the markers represent the simulation data points while the solid lines represent the theoretical approximations. Note that the light shaded band appearing in the $u_{dx}$ characteristics are the markers representing simulation data points over a wide range of St. Here the simulation data are taken for $a/H = 0.4375$ and $Ca = 0.286$.

Now, starting from Eq. (18), the oscillating flow condition $(\varepsilon_1 = 0, \varepsilon_2 = 1)$ implies

$$u_{dx}^{(0)} \propto \sin(t\,\text{St}) \tag{19}$$

subsequently we have

$$x_d \propto \int \sin(t\,\text{St})dt \quad \Rightarrow \quad x_d \propto \frac{1}{\text{St}}\cos(t\,\text{St}). \tag{20}$$

Fig. 7a shows that the present scaling notion (Eq. (19) and (20)) gives fairly good agreement with the simulation results, at least for describing the time complexity of $u_{dx}$. In tune with Eq. (19), we have $\text{St}_{dx} = \text{St}$ (Fig. 7b), with $\text{St}_{dx}$ being the frequency of oscillation in $u_{dx}$. Subsequently, from Eq. (20) we obtain the relation $(x_d)_{amp} \propto \text{St}^{-1}$ (Fig. 7c), with $(x_d)_{amp}$ being the amplitude of $x_d$. This inverse relationship between $(x_d)_{amp}$ and St explains the difference in the $x_d - \theta$ patterns at different St, as already emphasized in Fig. 5. Note that here the simulation data are taken for $a/H = 0.4375$ and $\text{Ca} = 0.286$, to highlight the features for confined droplets. However, from all the plots of Fig. 7, we note the departure of the simulation results from the present scaling notion (equations (19) and (20)). This may be attributed to the increased difference between droplet response time scale with the time scale of alteration in the flow field, at higher St.

### D. Effect of inertia due to imposed time complexity: the $O(\text{Re})$ problem

The governing equations for the $O(\text{Re})$ problem are

$$\left.\begin{aligned}
(a)\ &\nabla \cdot \mathbf{u}^{(\text{Re})} = 0 \text{ and } \nabla \cdot \mathbf{T}^{(\text{Re})} = \frac{\partial \mathbf{u}^{(0)}}{\partial t} + \mathbf{u}^{(0)} \cdot \nabla \mathbf{u}^{(0)} \\
&\text{with } \mathbf{T}^{(\text{Re})} = -p^{(\text{Re})}\mathbf{I} + \left(\nabla \mathbf{u}^{(\text{Re})} + \nabla \mathbf{u}^{(\text{Re})T}\right), \\
(b)\ &\nabla \cdot \bar{\mathbf{u}}^{(\text{Re})} = 0 \text{ and } \nabla \cdot \bar{\mathbf{T}}^{(\text{Re})} = \frac{\partial \bar{\mathbf{u}}^{(0)}}{\partial t} + \bar{\mathbf{u}}^{(0)} \cdot \nabla \bar{\mathbf{u}}^{(0)} \\
&\text{with } \bar{\mathbf{T}}^{(\text{Re})} = -\lambda \bar{p}^{(\text{Re})}\mathbf{I} + \lambda \left(\nabla \bar{\mathbf{u}}^{(\text{Re})} + \nabla \bar{\mathbf{u}}^{(\text{Re})T}\right), \\
(c)\ &\nabla \cdot \mathbf{u}_\infty^{(\text{Re})} = 0 \text{ and } \nabla \cdot \mathbf{T}_\infty^{(\text{Re})} = \frac{\partial \bar{\mathbf{u}}_\infty^{(0)}}{\partial t} + \bar{\mathbf{u}}_\infty^{(0)} \cdot \nabla \bar{\mathbf{u}}_\infty^{(0)} \\
&\text{with } \mathbf{T}_\infty^{(\text{Re})} = -p_\infty^{(\text{Re})}\mathbf{I} + \left(\nabla \mathbf{u}_\infty^{(\text{Re})} + \nabla \mathbf{u}_\infty^{(\text{Re})T}\right).
\end{aligned}\right\} \quad (21)$$

with the boundary and interfacial conditions

$$\left.\begin{aligned}
(i)\ &\mathbf{u}^{(\text{Re})} \to \mathbf{u}_\infty^{(\text{Re})} \text{ as } r \to \infty, \\
(ii)\ &\mathbf{u}^{(\text{Re})} = -\mathbf{u}_d^{(\text{Re})} \text{ on the walls}, \\
(iii)\ &\mathbf{u}^{(\text{Re})} = \bar{\mathbf{u}}^{(\text{Re})} \text{ at } r = 1, \\
(iv)\ &\mathbf{u}^{(\text{Re})} \cdot \hat{\mathbf{e}}_r = \bar{\mathbf{u}}^{(\text{Re})} \cdot \hat{\mathbf{e}}_r = 0 \text{ at } r = 1, \text{ and} \\
(v)\ &\mathbf{T}^{(\text{Re})} \cdot \hat{\mathbf{e}}_r - \bar{\mathbf{T}}^{(\text{Re})} \cdot \hat{\mathbf{e}}_r = -\left[2f^{(\text{ReCa})} + \nabla^2 f^{(\text{ReCa})}\right]\hat{\mathbf{e}}_r \text{ at } r = 1.
\end{aligned}\right\} \quad (22)$$

Subsequently, the reciprocal identity (Eq. (10)) transforms to

$$\int_{S_d} \left[\left(\mathbf{T}^{(\text{Re})} - \bar{\mathbf{T}}^{(\text{Re})}\right) \cdot \mathbf{u}_c - \left(\mathbf{T}_c - \bar{\mathbf{T}}_c\right) \cdot \mathbf{u}^{(\text{Re})} \right.$$
$$\left. - \bar{\mathbf{T}}_c \cdot \left(\mathbf{u}^{(\text{Re})} - \bar{\mathbf{u}}^{(\text{Re})}\right) - \mathbf{T}_\infty^{(\text{Re})} \cdot \mathbf{u}_c + \mathbf{T}_c \cdot \mathbf{u}_\infty^{(\text{Re})}\right] \cdot \hat{\mathbf{e}}_r dS =$$

$$-\left\{\int_\Omega\left[\left(\frac{\partial \mathbf{u}^{(0)}}{\partial t}+\mathbf{u}^{(0)}\cdot\nabla\mathbf{u}^{(0)}\right)-\left(\frac{\partial \mathbf{u}_\infty^{(0)}}{\partial t}+\mathbf{u}_\infty^{(0)}\cdot\nabla\mathbf{u}_\infty^{(0)}\right)\right]\cdot\mathbf{u}_c d\Omega\right.$$
$$\left.-\int_\Omega\left[\frac{\partial \bar{\mathbf{u}}^{(0)}}{\partial t}+\bar{\mathbf{u}}^{(0)}\cdot\nabla\bar{\mathbf{u}}^{(0)}\right]\cdot\bar{\mathbf{u}}_c d\Omega\right\}. \tag{23}$$

From the reciprocal identity (Eq. (23)), it is evident that we require the $O(\text{Re})$ solution of the imposed flow field. This degenerates to the form

$$\mathbf{u}_\infty^{(\text{Re})}=\left[\frac{\alpha}{2}\left(\psi^2+\alpha'\right)+\frac{\beta}{6}\left(\psi^3+\beta'\right)+\frac{\gamma}{12}\left(\psi^4+\gamma'\right)\right]\dot{\Gamma}(t)\hat{\mathbf{e}}_\xi-\frac{y_d}{2a/H}\dot{\mathbf{u}}_d^{(0)}-\mathbf{u}_d^{(\text{Re})} \tag{24}$$

with $\alpha'=s\left[(a/H)^{-1}-y_d\right]$, $\beta'=2y_d\left[(a/H)^2-y_d^2\right]/3$, $\gamma'=6y_d\left[(a/H)^{-3}-y_d^3\right]$ and the dots represent the time derivative. Additionally, we require the solution of the complementary problem. Without any loss of generality we can adopt the solution from Ref. [11].

With the above mentioned arguments through Eqs. (21)-(24), we eventually end up into the relation

$$u_{dy}^{(\text{Re})}=\chi_1' u_{dx}^{(\text{Re})}+\chi_2' u_{dx}^{(0)}+\chi_3' \dot{u}_{dx}^{(0)}+\chi_4'\dot{\Gamma}(t)+\chi_5' \tag{25}$$

where the coefficients $\chi'$ can be obtained from the reciprocal identity (Eq. (23)), using the $O(1)$ solutions for the entire problem and the $O(\text{Re})$ solution of the imposed flow field. Note that here we are interested in unveiling the time complex features. Thus, the exact forms of $\chi'$s are not important here. In expressing $u_{dy}^{(\text{Re})}$ through Eq. (25), all quantities are known except $u_{dx}^{(\text{Re})}$. Without any loss of generality, we may invoke the matching condition $\mathbf{u}^{(\text{Re})}\to\mathbf{u}_\infty^{(\text{Re})}$ as $r\to\infty$ to approximate the temporal nature of $\mathbf{u}^{(\text{Re})}$. The complete description of $\mathbf{u}^{(\text{Re})}$, however, requires the $O(\text{Re})$ level solution of the problem with proper accounting of the overlapping of the outer and the inner region. Restricting our attention to the analysis of time complexity, thus, from the force estimation we expect the nature of $u_{dx}^{(\text{Re})}$ is of the form

$$u_{dx}^{(\text{Re})}=\Xi_1+\Xi_2\dot{\Gamma}(t)+\Xi_3\dot{u}_{dx}^{(0)}. \tag{26}$$

with $\Xi$s being the coefficients.

Now, using equations (18) and (26) into Eq. (25), we obtain

$$u_{dy}^{(\text{Re})}=\chi_1+\chi_2\Gamma(t)+\chi_3\dot{\Gamma}(t) \tag{27}$$

with $\chi$ s being the new constants. Eq. (27) can be considered as the generic scaling estimation of $u_{dy}^{(Re)}$ to describe its time complexity.

### E. Effect of deformation: the $O(\text{Ca})$ problem

The governing equations for the $O(\text{Ca})$ problem are

$$\left.\begin{aligned}
&(a)\,\nabla\cdot\mathbf{u}^{(\text{Ca})} = 0 \text{ and } \nabla\cdot\mathbf{T}^{(\text{Ca})} = 0 \text{ with } \mathbf{T}^{(\text{Ca})} = -p^{(\text{Ca})}\mathbf{I} + \left(\nabla\mathbf{u}^{(\text{Ca})} + \nabla\mathbf{u}^{(\text{Ca})T}\right),\\
&(b)\,\nabla\cdot\bar{\mathbf{u}}^{(\text{Ca})} = 0 \text{ and } \nabla\cdot\bar{\mathbf{T}}^{(\text{Ca})} = 0 \text{ with } \bar{\mathbf{T}}^{(\text{Ca})} = -\lambda\bar{p}^{(\text{Ca})}\mathbf{I} + \lambda\left(\nabla\bar{\mathbf{u}}^{(\text{Ca})} + \nabla\bar{\mathbf{u}}^{(\text{Ca})T}\right),\\
&(c)\,\nabla\cdot\mathbf{u}_\infty^{(\text{Ca})} = 0 \text{ and } \nabla\cdot\mathbf{T}_\infty^{(\text{Ca})} = 0 \text{ with } \mathbf{T}_\infty^{(\text{Ca})} = -p_\infty^{(\text{Ca})}\mathbf{I} + \left(\nabla\mathbf{u}_\infty^{(\text{Ca})} + \nabla\mathbf{u}_\infty^{(\text{Ca})T}\right),
\end{aligned}\right\} \quad (28)$$

with the boundary and interfacial conditions

$$\left.\begin{aligned}
&(i)\,\mathbf{u}^{(\text{Ca})} \to \mathbf{u}_\infty^{(\text{Ca})} \text{ as } r \to \infty,\\
&(ii)\,\mathbf{u}^{(\text{Ca})} = -\mathbf{u}_d^{(\text{Ca})} \text{ on the walls,}\\
&(iii)\,\mathbf{u}^{(\text{Ca})} + f^{(\text{Ca})}\frac{\partial \mathbf{u}^{(0)}}{\partial r} = \bar{\mathbf{u}}^{(\text{Ca})} + f^{(\text{Ca})}\frac{\partial \bar{\mathbf{u}}^{(0)}}{\partial r} \text{ at } r=1,\\
&(iv)\,\left[\mathbf{u}^{(\text{Ca})} + f^{(\text{Ca})}\frac{\partial \mathbf{u}^{(0)}}{\partial r}\right]\cdot\hat{\mathbf{e}}_r - \mathbf{u}^{(0)}\cdot\nabla f^{(Ca)} = \left[\bar{\mathbf{u}}^{(\text{Ca})} + f^{(\text{Ca})}\frac{\partial \bar{\mathbf{u}}^{(0)}}{\partial r}\right]\cdot\hat{\mathbf{e}}_r - \bar{\mathbf{u}}^{(0)}\cdot\nabla f^{(Ca)} = 0 \text{ at } r=1, \text{ and}\\
&(v)\,\left[\mathbf{T}^{(\text{Ca})} + f^{(\text{Ca})}\frac{\partial \mathbf{T}^{(0)}}{\partial r}\right]\cdot\hat{\mathbf{e}}_r - \left[\bar{\mathbf{T}}^{(\text{Ca})} + f^{(\text{Ca})}\frac{\partial \bar{\mathbf{T}}^{(0)}}{\partial r}\right]\cdot\hat{\mathbf{e}}_r - \left[\mathbf{T}^{(0)} - \bar{\mathbf{T}}^{(0)}\right]\cdot\nabla f^{(\text{Ca})}\\
&\qquad = \left[2f^{(\text{Ca})} + \nabla^2 f^{(\text{Ca})}\right]\nabla f^{(\text{Ca})} - \left[2f^{(\text{CaCa})} - 2f^{(\text{Ca})}f^{(\text{Ca})} + \nabla^2 f^{(\text{CaCa})}\right]\hat{\mathbf{e}}_r\\
&\qquad\qquad\qquad\qquad\qquad\qquad\qquad\qquad\qquad\qquad\qquad\qquad\qquad \text{at } r=1.
\end{aligned}\right\}$$
(29)

Subsequently, the reciprocal identity (Eq. (10)) transforms to

$$\int_{S_d}\left[\left(\mathbf{T}^{(\text{Ca})} - \bar{\mathbf{T}}^{(\text{Ca})}\right)\cdot\mathbf{u}_c - \left(\mathbf{T}_c - \bar{\mathbf{T}}_c\right)\cdot\mathbf{u}^{(\text{Ca})} - \bar{\mathbf{T}}_c\cdot\left(\mathbf{u}^{(\text{Ca})} - \bar{\mathbf{u}}^{(\text{Ca})}\right)\right.$$
$$\left. - \mathbf{T}_\infty^{(\text{Ca})}\cdot\mathbf{u}_c + \mathbf{T}_c\cdot\mathbf{u}_\infty^{(\text{Ca})}\right]\cdot\hat{\mathbf{e}}_r\, dS = 0. \quad (30)$$

Now we approximate different integrals using the matching conditions from Eq. (29) [11]:

$$(i)\int_{S_d}\left[\left(\mathbf{T}^{(\text{Ca})} - \bar{\mathbf{T}}^{(\text{Ca})}\right)\cdot\mathbf{u}_c\cdot\hat{\mathbf{e}}_r\, dS = \int_{S_d}\left[-f^{(\text{Ca})}\frac{\partial}{\partial r}\left(\mathbf{T}^{(0)} - \bar{\mathbf{T}}^{(0)}\right)\cdot\hat{\mathbf{e}}_r\right.$$
$$\left. + \left(\mathbf{T}^{(0)} - \bar{\mathbf{T}}^{(0)}\right)\cdot\nabla f^{(\text{Ca})} + \left(2f^{(\text{Ca})} + \nabla^2 f^{(\text{Ca})}\right)\nabla f^{(\text{Ca})}\right]\cdot\mathbf{u}_c\, dS$$

$$(ii) -\int_{S_d} \left(\mathbf{T}_c - \bar{\mathbf{T}}_c\right) \cdot \mathbf{u}^{(Ca)} \cdot \hat{\mathbf{e}}_r dS = \int_{S_d} \left(\mathbf{T}_c - \bar{\mathbf{T}}_c\right) : \hat{\mathbf{e}}_r \hat{\mathbf{e}}_r$$

$$\left[ f^{(Ca)} \frac{\partial \mathbf{u}^{(0)}}{\partial r} \cdot \hat{\mathbf{e}}_r - \mathbf{u}^{(0)} \cdot \nabla f^{(Ca)} \right] dS$$

$$(iii) -\int_{S_d} \bar{\mathbf{T}}_c \cdot \left(\mathbf{u}^{(Ca)} - \bar{\mathbf{u}}^{(Ca)}\right) \cdot \hat{\mathbf{e}}_r dS = \int_{S_d} \bar{\mathbf{T}}_c \cdot \hat{\mathbf{e}}_r \cdot \left[ f^{(Ca)} \frac{\partial}{\partial r} \left(\mathbf{u}^{(0)} - \bar{\mathbf{u}}^{(0)}\right) \right] dS \tag{31}$$

$$(iv) \int_{S_d} \left[ -\mathbf{T}_\infty^{(Ca)} \cdot \mathbf{u}_c + \mathbf{T}_c \cdot \mathbf{u}_\infty^{(Ca)} \right] \cdot \hat{\mathbf{e}}_r dS = -2\pi \frac{2+3\lambda}{1+\lambda} (1 + J_1 + J_4) u_{dy}^{(Ca)}$$

where $J_1$ and $J_4$ have the same meaning as in Ref. [11].

From Eq. (31), it is evident that for estimating $u_{dy}^{(Ca)}$, we need to have the $O(1)$ solution and estimation of $f^{(Ca)}$. The later can be estimated using the matching condition from Eq. (16)

$$\mathbf{T}^{(0)} \cdot \hat{\mathbf{e}}_r - \bar{\mathbf{T}}^{(0)} \cdot \hat{\mathbf{e}}_r = -\left[ 2f^{(Ca)} + \nabla^2 f^{(Ca)} \right] \hat{\mathbf{e}}_r \text{ at } r = 1. \tag{32}$$

Now, introducing the $O(1)$ solution, it is sufficient to consider $\left(\mathbf{T}^{(0)} \cdot \hat{\mathbf{e}}_r - \bar{\mathbf{T}}^{(0)} \cdot \hat{\mathbf{e}}_r\right) \propto \Gamma(t)$. In tune with the matching condition (Eq. (32)), thus, we can approximate that $f^{(Ca)} \propto \Gamma(t)$. If we consider the integrals in Eq. (31), they are of the form $\int_{S_d} f^{(Ca)} \cdot \mathbf{T}^{(0)} \cdot \hat{\mathbf{e}}_r$ or $\int_{S_d} f^{(Ca)} \cdot \mathbf{u}^{(0)} \cdot \hat{\mathbf{e}}_r$. Invoking the estimation $f^{(Ca)} \propto \Gamma(t)$, $\mathbf{u}^{(0)} \propto \Gamma(t)$ and $\mathbf{T}^{(0)} \propto \Gamma(t)$, thus, we can represent $u_{dy}^{(Ca)}$ as

$$u_{dy}^{(Ca)} = \Pi_1 + \Pi_2 \Gamma(t)^2 \tag{33}$$

with $\Pi$s being the coefficients.

### F. Combined influence of inertia and deformation on the time complexity of the lateral movement

To introduce the combined influence of inertia and deformation, to the leading order approximation, it is sufficient to recall the $O(\text{Re})$ and $O(\text{Ca})$ approximation of $u_{dy}$, as

$$\begin{aligned} u_{dy} &= \text{Re}\, u_{dy}^{(Re)} + \text{Ca}\, u_{dy}^{(Ca)} + \ldots \\ &= \text{Re} \left[ \chi_1 + \chi_2 \Gamma(t) + \chi_3 \dot{\Gamma}(t) \right] + \text{Ca} \left[ \Pi_1 + \Pi_2 \Gamma(t)^2 \right] + \ldots \end{aligned} \tag{34}$$

Introducing $\Gamma(t) = \varepsilon_1 + \varepsilon_2 \sin(t\,\text{St})$, Eq. (34) can be given by

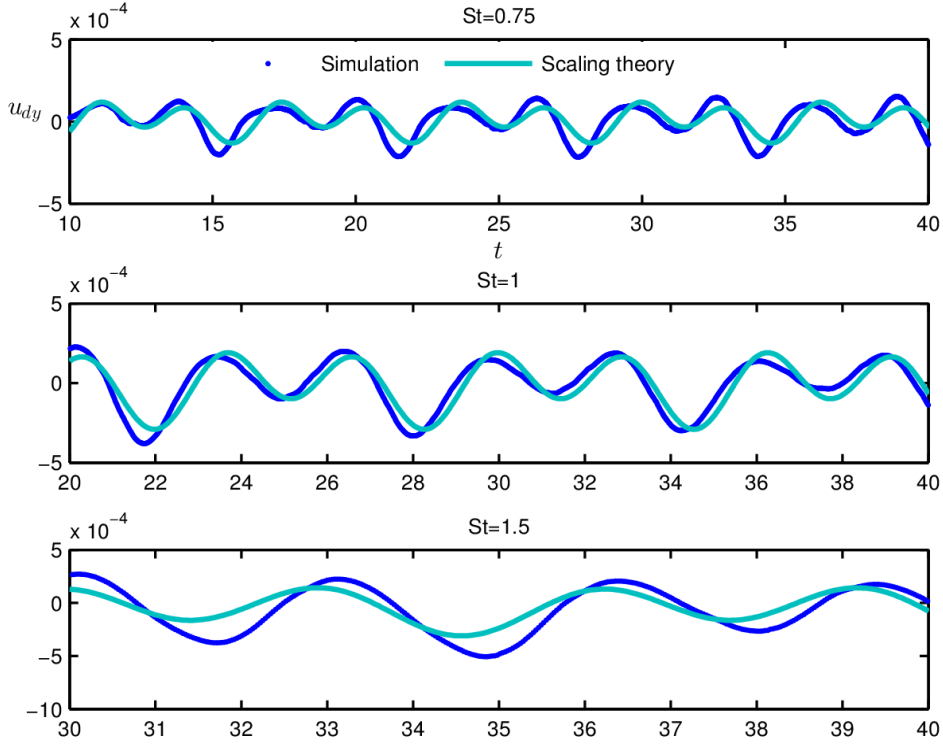

FIG. 8. (Color online) Comparison of the present scaling estimation of the time complexity of $u_{dy}$ with the simulation data points different $St$. Here, the simulation data are taken for $a/H = 0.4375$ and $Ca = 0.286$.

$$u_{dy} = \mathrm{Re}\left[\chi_1 + \chi_2\{\varepsilon_1 + \varepsilon_2 \sin(tSt)\} + \chi_3 \varepsilon_2 St \cos(tSt)\right] \\ + Ca\left[\Pi_1 + \Pi_2\{\varepsilon_1^2 + \varepsilon_2^2 \sin^2(tSt) + 2\varepsilon_1\varepsilon_2 \sin(tSt)\}\right] + \ldots \qquad (35)$$

For an oscillating flow condition $(\varepsilon_1 = 0, \varepsilon_2 = 1)$, Eq. (35) yields

$$u_{dy} = \mathrm{Re}\left[\chi_1 + \chi_2 \sin(tSt) + \chi_3 St \cos(tSt)\right] \\ + Ca\left[\Pi_1 + \Pi_2 \sin^2(tSt)\right] + \ldots \qquad (36)$$

Starting from Eq. (36), $u_{dy}$ can be given by a scaling relationship of the form

$$u_{dy} = c_1 + c_2 \sin(tSt) + c_3 \sin^2(tSt) + c_4 St \cos(tSt), \qquad (37)$$

with $c$s being the coefficients. Eq. (37) describes a generic scaling estimation for the time complexity of the lateral migration speed due to the imposed oscillation in the flow field. Fig. 8 compares this scaling relationship with the present simulation results at different $St$. Note that here the simulation data are taken for $a/H = 0.4375$ and $Ca = 0.286$, to highlight the features for confined droplets. From the figure, it is evident that our theoretical analysis is sufficient to bring out the time complexity of dynamical features of a confined droplet in an oscillating flow field.

## V. CONCLUDING REMARKS

The dynamics of a confined droplet in a time complex flow field is a consequence of the combined influence of the droplet deformation and the intrinsic inertia. However, at low frequency of imposed oscillation, the streamwise motion characteristics of the droplet can be approximated from spherical droplet consideration under quasi-steady Stokes flow approximation in the low Reynolds number limit. The cross-stream migration characteristics, on the other hand, are decided by the combined contribution of deformation and inertia at all circumstances. Moreover, the wall induced lift force is also of immense significance. As a result, the droplet executes transverse oscillation. To gain deeper insight, a theoretical estimation of the droplet trajectory may be provided based on the reciprocal theorem. We have shown that the consequent estimations for the time scales of the droplet migration characteristics corroborate the findings from the numerical simulation results obtained through phase-field formalism. These findings can be of potential importance for many microfluidics based applications, including micro-reactors for mixing augmentation and reaction analysis.